\documentclass[fleqn,usenatbib]{mnras}

\usepackage{newtxtext,newtxmath}
\usepackage[T1]{fontenc}
\usepackage{ae,aecompl}

\usepackage{graphicx}	% Including figure files
\usepackage{amsmath}	% Advanced maths commands
\usepackage{wasysym}
\usepackage{xspace}
\usepackage{float}
\usepackage[normalem]{ulem}
\usepackage{wrapfig}
\usepackage{xcolor}
\title[KDG~64: a dSph/UDG satellite of M~81]{KDG~64: a large dwarf spheroidal or a small ultradiffuse satellite of Messier~81}

\author[A. V. Afanasiev et al.]{Anton V. Afanasiev,$^{1,2}$\thanks{E-mail: anton.afanasiev@voxastro.org}
Igor V. Chilingarian,$^{3,2}$\thanks{E-mail: igor.chilingarian@cfa.harvard.edu; chil@sai.msu.ru}
Kirill A. Grishin,$^{1,2}$ \newauthor
Dmitry Makarov,$^{4}$ 
Lidia Makarova,$^{4}$
Daniel Fabricant,$^{3}$
Nelson Caldwell,$^{3}$ \newauthor
and Sean Moran$^{3}$
\\
$^{1}$Universit\'{e} Paris Cit\'{e}, CNRS, Astroparticule et Cosmologie, F-75013 Paris, France\\
$^{2}$Sternberg Astronomical Institute, M.~V.~Lomonosov Moscow State University, Universitetsky prospect 13, Moscow, 119234, Russia\\
$^{3}$Center for Astrophysics --- Harvard and Smithsonian, 60 Garden St. MS09, Cambridge, MA 02138, USA\\
$^{4}$Special Astrophysical Observatory of RAS, Nizhnij Arkhyz 369167, Karachai-Cherkessian Republic, Russia\\
}
\date{Accepted 2023 February 10. Received 2023 February 9; in original form 2022 June 12}

\pubyear{2023}
\volume{520}
\pagerange{6312-6321}

\begin{document}

\label{firstpage}
\maketitle

\begin{abstract}
Low-mass early-type galaxies, including dwarf spheroidals (dSph) and brighter dwarf ellipticals (dE), dominate the galaxy population in groups and clusters. Recently, an additional early-type population of more extended ultra-diffuse galaxies (UDGs) has been identified, sparking a discussion on the potential morphological and evolutionary connections between the three classifications. Here we present the first measurements of spatially resolved stellar kinematics from deep integrated-light spectra of KDG~64 (UGC~5442), a large dSph galaxy in the M~81 group. From these data we infer stellar population properties and dark matter halo parameters using Jeans dynamical modelling. We find an old, metal-poor stellar population with no young stars and a dark matter mass fraction of $\sim 90$~per~cent within the half-light radius. These properties and the position of KDG~64 on the Fundamental Plane indicate that it is a local analogue of smaller UDGs in the Coma and Virgo clusters and is probably a transitional dSph-UDG object. Its evolutionary path cannot be uniquely established from the existing data, but we argue that supernovae feedback and tidal heating played key roles in shaping KDG~64.
\end{abstract}


%%%%%%%%%%%%%%%%%%%%%%%%%%%%%%%%%%%%%%%%%%%%%%%%%%

%%%%%%%%%%%%%%%%% BODY OF PAPER %%%%%%%%%%%%%%%%%%

\section{Introduction}

Low-luminosity early-type galaxies with no current star formation are the numerically dominant population in galaxy clusters \citep{SB84,FS88}. These galaxies are commonly called dwarf ellipticals (dE) or dwarf lenticulars (dS0) and typically have stellar masses $M_* \sim 10^8 - 10^9 M_{\odot}$ and effective radii $0.5-2$~kpc.  Galaxy groups, on the other hand, are numerically dominated by yet fainter dwarf spheroidal galaxies (dSph) similar to dE in size (half-light radius $R_e\sim0.5-2$~kpc) and morphology but with $10-100$ times fewer stars \citep{2003AJ....125.1926G}. \citet{SB84} found extended low-surface brightness galaxies with larger radii ($R_e = 1.5 - 4.5$~kpc) and stellar masses similar or smaller than dEs.  \citet[][]{2015ApJ...798L..45V} proposed that these ultra-diffuse galaxies (UDGs) constitute a distinct galaxy class. UDGs were found in large quantities first in the Coma cluster and later in other clusters and groups \citep{2015ApJ...813L..15M,2015ApJ...809L..21M,2017ApJ...839L..17J,2017MNRAS.470.1512W,2021ApJS..257...60Z}. 

How these morphological classes relate to each other and the evolutionary processes leading to their structural differences remain uncertain \citep{2018RNAAS...2a..43C}. UDGs are the most puzzling because they share characteristics with both with dE and dSph \citep{Chilingarian+19}. Understanding the origin of UDGs is a crucial step in our studies of the evolutionary history of early-type dwarf galaxies. As of now, the UDG class remains ambiguously defined because different teams used different size and surface brightness limits to separate it from dEs and dSphs, e.g. \citet{2015ApJ...798L..45V} applied a half-light radius cut of $R_e>1.5$~kpc from shallow CFHT images, and to recover their population from much deeper Subaru images, \citet{Koda15} had to reduce the $R_e$ threshold to 0.7~kpc.

The low UDG surface brightness makes spectroscopy difficult \citep{2021NatAs...5.1308G}, and we have measurements of stellar population parameters and stellar velocity dispersions for only a handful of UDGs  \citep{2016ApJ...819L..20B,2016ApJ...828L...6V,Danieli+19, Chilingarian+19,2019ApJ...880...91V,2020MNRAS.495.2582G,2021MNRAS.502.3144G}.  The internal dynamics and dark matter content of UDGs remain poorly constrained. 

The nearest known galaxy classified as a non-starforming UDG outside the Local Group is 13~Mpc distant \citep{2019ApJ...880L..11M}. At that distance only the brightest RGB stars are detectable, using HST, thus detailed stellar population analysis is impossible. Finding UDG-like candidates in the nearest groups would help us understand UDG formation and evolution. The largest quiescent dwarf satellite of the Milky Way with the stellar mass comparable to that of UDGs in nearby galaxy clusters, Fornax dSph, clearly cannot be considered as a UDG, because of its relatively small 700~pc effective radius and $\sim 23$ mag~arcsec$^{-2}$ average surface brightness \citep{2006ApJ...638..725Z}. The Local Group contains a number of extended galaxies with the half-light radii as large as $r=3$~kpc \citep{2012AJ....144....4M} determined from the analysis of their resolved red giant populations to equivalent surface brightnesses of $\mu_g>30$~mag~arcsec$^{-2}$, however (i) such faint systems cannot be detected \emph{en masse} in integrated light beyond the Local Volume, and (ii) their stellar masses are orders of magnitude smaller than those of UDGs in the Coma cluster identified by \citet{2015ApJ...798L..45V} and then \citet{Koda15}.

\begin{figure}
\vskip-6mm
    \centering
    \includegraphics[width=0.97\hsize]{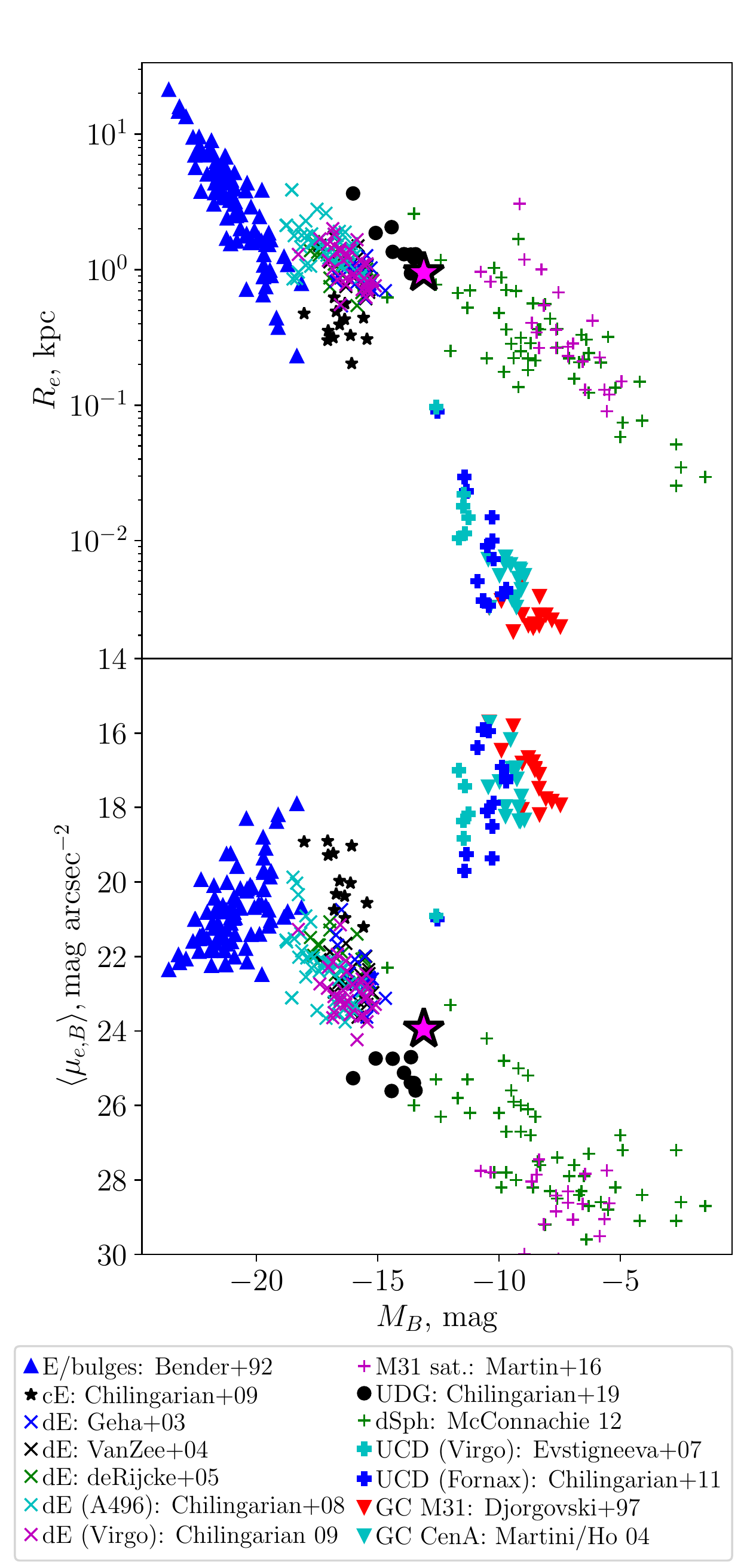}
    \caption{Size -- luminosity and mean effective surface brightness -- luminosity relations of quiescent early-type galaxies and compact stellar systems. KDG~64 is marked by a pink star. The data sources are shown in the legend.} 
    \label{fig:plot_sbrel}
\end{figure}

The M~81 group has 4 dwarf quiescent galaxies with r$_e$ $\geq$ 1~kpc \citep{2019ApJ...884..128O} but one \citep[F8D1,][]{1998AJ....115..535C} is located behind Galactic cirrus, another \citep[IKN,][]{2004AJ....127.2031K} lies close to a 9-th~mag star, and a third \citep[KDG~61,][]{Makarova10} is located very near M~81 with projected foreground star-forming regions and likely tidal disturbance. The remaining dSph, KDG~64, is an excellent candidate for deep integrated-light spectroscopy. KDG~64 is one of the largest known dSph, and it borders the UDG and dE regimes in size and surface brightness. In Fig.~\ref{fig:plot_sbrel} we show a compilation of the literature data for the structural parameters (luminosity, half-light radius, mean surface brightness within $R_e$; c.f. \citealp{Kormendy77}) for dwarf and giant early-type galaxies and compact stellar systems. KDG~64 marked with a large purple star lies between the loci of dSphs and dEs at end edge of the area populated by Coma cluster UDGs.

\nocite{2016ApJ...833..167M}

KDG~64 is interesting because it borders multiple dwarf galaxy sub-classes. Its distance of $3.73$~Mpc \citep{2016AJ....152...50T} enables HST photometry of individual stars but precludes obtaining individual stellar spectra. KDG~64's angular size, surface density and distance places sufficient stars located within a spectrograph slit to measure the average stellar velocity dispersion ($\sigma$) along the line of sight.  The stochastic influence of individual stellar velocities is insignificant \citep{1992ApJ...400..510D}.  Our pilot study of KDG~64 demonstrates our ability to measure internal kinematics, to determine the stellar population, and to infer the dark matter content through dynamical modelling. A similar approach can be applied to future studies of dwarf galaxies with a wider range of masses, luminosities, surface brightnesses and dark matter properties. \vskip -5mm

\begin{figure*}
\centering
\includegraphics[width=\hsize]{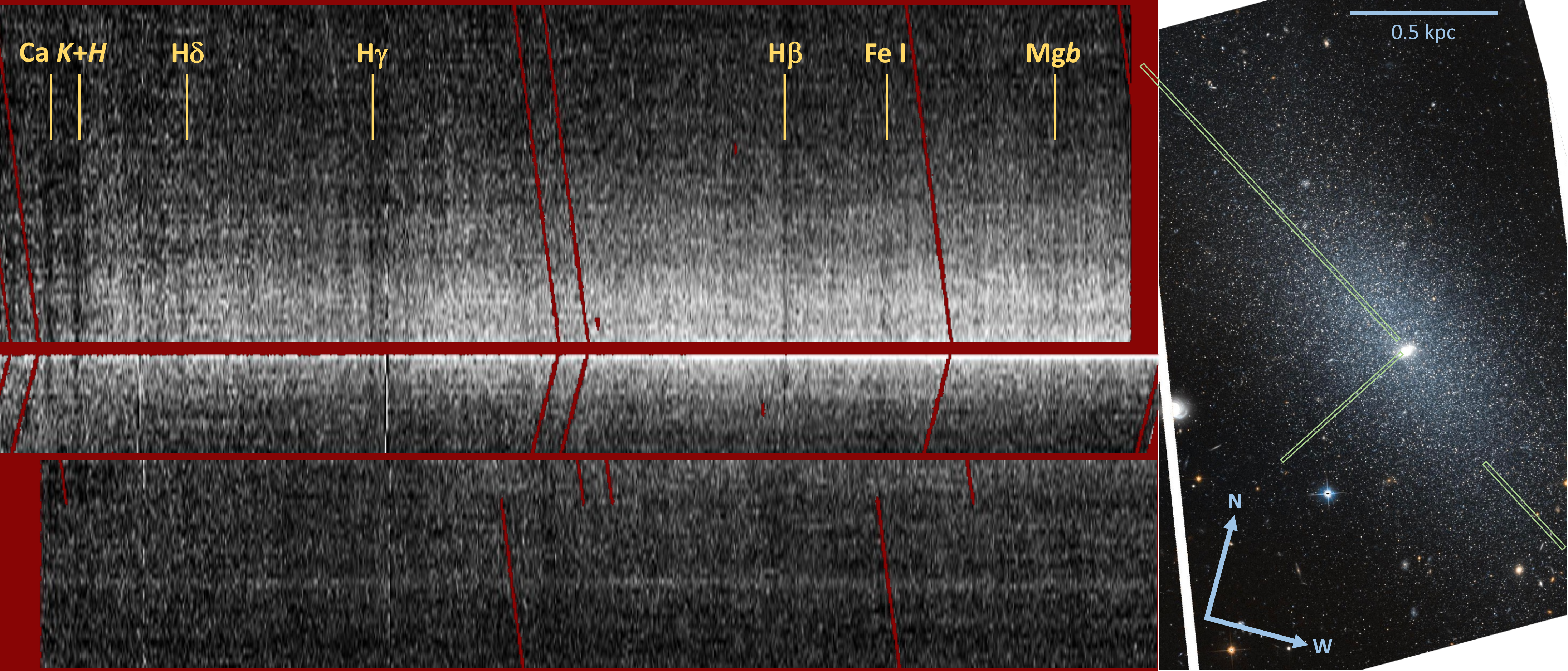}
  \caption{\textbf{Left:} long slit spectra along galaxy major and minor axes. \textbf{Right:} slit positions overlaid on an HST false-color (\textit{F606W/F814W}) image. \label{fig:mask}}
\end{figure*}

\section{New and archival observations}

We observed KDG~64 with Binospec, a multi-object spectrograph at the 6.5-m MMT  \citep{2019PASP..131g5004F}. We designed a single slit mask with three primary slits for KDG~64: two 50~arcsec-long slits along the major axis and one 20~arcsec-long slit along the minor axis. The major and minor axis slits are tilted by 35~deg and 55~deg to the direction of the dispersion, respectively. To prevent spectra from overlapping, the second major axis slit begins 28.5~arcsec from the galaxy centre. We excluded a background galaxy located near the geometric centre of KDG~64. The mask PA value was set to +165~deg (see Fig.~\ref{fig:mask}).  

The spectra were obtained on two consecutive nights: November 6\&7, 2018.  We used Binospec's 1000~gpm gratings with wavelength coverage 3,760~\AA\ -- 5,300~\AA\ and $R=3,750 - 4,900$ ($\sigma_{\mathrm{inst}} = 34 - 26$~km~s$^{-1}$). The total integration time of 3h~20~min was split into 20~min-long exposures. The observations were conducted with seeing 0.9--1.3~arcsec and good transparency during dark time. We took arc lamp and internal flat field frames at night time and high signal-to-noise sky flats during the day to characterize the spectral resolution across the field and wavelength range.

We reduced the data with the Binospec pipeline\footnote{\url{https://bitbucket.org/chil_sai/binospec/src}} \citep{2019PASP..131g5005K} that produces flux calibrated, sky-subtracted, rectified, and wavelength calibrated 2D long-slit images (plus flux error frames) for each slitlet. The image scale is 0.24~arscec~pix$^{-1}$ along the slit and the wavelength sampling is 0.38~\AA~pix$^{-1}$. We used non-local sky subtraction (i.e. a global sky model computed from all ``empty'' regions in all slits in the mask) optimized for low surface brightness targets.

HST Advanced Camera for Surveys (ACS) archival images (obtained for a TRGB distance measurement) resolve KDG~64 into individual stars (PID~9884, PI: Armandroff) in the \textit{F606W} and \textit{F814W} (Cousins $Ic$) bands.  We use archival $R$- and $V$-band images from the 2.1m KPNO telescope (PID~0117) for low-resolution 2-D surface photometry (300~s exposure time per band and 0.3~arcsec~pix$^{-1}$ scale).  We construct a broadband spectral energy distribution (SED) with aperture photometry from GALEX  (NUV and FUV bands), HST (\textit{F606W} and \textit{F814W} bands), KPNO (B, V, R bands) and Spitzer Space Telescope (\textit{IRAC1} $3.6~\mu m$ and \textit{IRAC2} $4.5~\mu m$).

\section{Data Analysis methods}

\begin{figure*}
    \centering
    \includegraphics[width=0.95\linewidth]{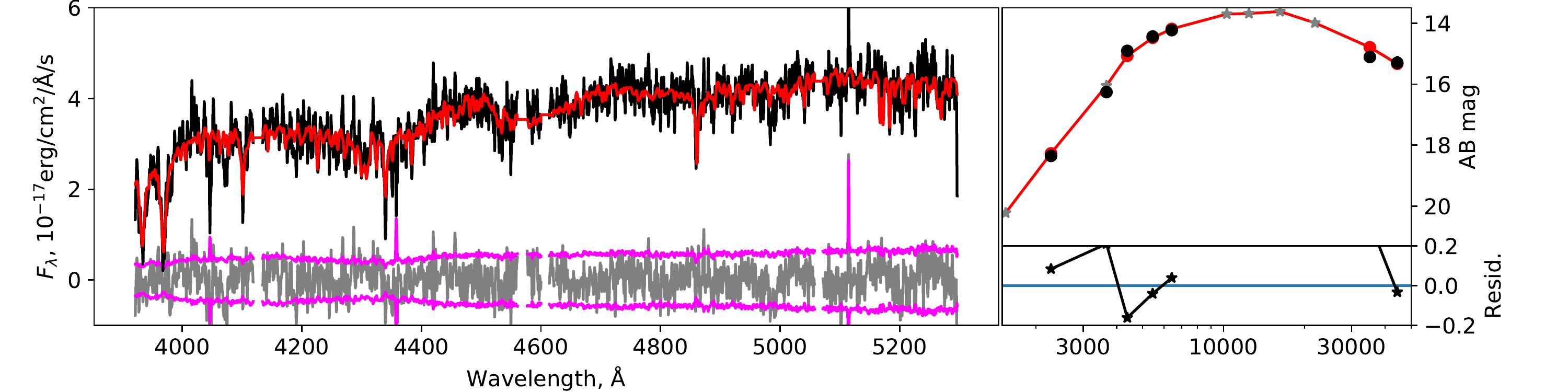}
    \caption{A Binospec spectrum of KDG~64 binned within 12~arcsec along the major axis on one side from the galaxy centre in $F_{\lambda}$ units (left panel, black); an observed SED in AB magnitudes (right panel, black); the best-fitting model spectra and SED (red), flux uncertainties (purple) and residuals (grey line and black asterisks). The SED fitting residuals are shown in the bottom-right panel.}
    \label{fig:nbursts_major_spec}
\end{figure*}

\subsection{Full spectrum fitting}

We analyse the global properties of the KDG~64 stellar populations by modelling the integrated spectrum from the inner 12~arcsec of the major axis slitlet (Fig.~\ref{fig:nbursts_major_spec}) and the multi-wavelength spectral energy distribution. We use the {\sc NBursts+phot} technique \citep{2012IAUS..284...26C} that computes the position of a local minimum of the $(1-w_\mathrm{phot}) \chi^2_\mathrm{spec} + w_\mathrm{phot}\chi^2_\mathrm{phot}$ statistic. Here $\chi^2_\mathrm{spec}$ and $\chi^2_\mathrm{phot}$ are the $\chi^2$ statistics for a spectrum and an SED respectively, and $w_\mathrm{phot}$ is the statistical weight of the residuals in the SED fitting.  We model the optical spectrum from spectral templates interpolated to the specific values of stellar age and metallicity, convolved with the Gaussian line of sight velocity distribution (LOSVD) and multiplied by a polynomial continuum. 

For global stellar population modelling we use {\sc Miles} \citep{2010MNRAS.404.1639V} simple stellar population (SSP) models computed for {\sc Padova} isochrones \citep{2000A&AS..141..371G}. The broadband SED is modelled using {\sc pegase.2}-based low-resolution templates \citep{1997A&A...326..950F,1999astro.ph.12179F}. A spectrum is degraded to a spectral resolution of 2.3~\AA. We simultaneously model the observed optical spectrum integrated in the inner 12~arcsec bin and spectral energy distribution from UV to NIR. A flux in each band is extracted from an image within the elliptical isophote ($b/a=0.5$) aligned with the major axis of $100~\mathrm{arcsec} \sim 1.4~R_e$ excluding the background galaxy near the centre of KDG~64 as well as several other background/foreground contaminants. To account for potential flux calibration errors in both the observed spectrum and {\sc Miles} SSPs, we include a 15th order polynomial continuum in the fitting procedure. 

To measure internal kinematics, we also analyse a full-resolution long-slit spectrum using the {\sc NBursts} technique \citep{CPSK07,CPSA07}, which implements a full spectrum fitting approach in the pixel space using a grid of intermediate-spectral resolution ($R=10,000$) simple stellar population (SSP) models computed with the {\sc pegase.hr} evolutionary synthesis package \citep{LeBorgne+04}. We convolve SSPs with the Binospec spectral line-spread function determined from the analysis of sky flats. As \citet{CCB08} demonstrated, the {\sc NBursts} algorithm can recover stellar velocity dispersions ($\sigma$) down to $\sigma_{\mathrm{inst}}/2$ at a signal-to-noise ratio of 5 per pixel. We use the restframe wavelength range between 3900\AA\ and 5200--5400\AA\ for the fitting procedure. See \citet{Chilingarian+08,Chilingarian09} for a detailed discussion regarding sensitivity, systematics, and limitations of the {\sc NBursts} full spectrum fitting. We allow a 15th order multiplicative polynomial continuum correction and a constant additive term to account for uncertainties in subtracting scattered light. 

For the reduced long-slit spectra, the median signal-to-noise ratio (SNR) per pixel in the wavelength range from 4,800--5,000~\AA\ reached its maximum value of 1.65 in the brightest pixel of the major axis, the closest to the galaxy centre. To reliably measure stellar velocity dispersion of 15~km~s$^{-1}$ for old ($10$~Gyr) metal-poor ([Fe/H]=$-1.5$) stellar population \citep{2020PASP..132f4503C}, we bin the Binospec dataset along the slit increase the SNR per spatial bin. We end up with a total of 15 spatial bins along the major axis and 4 bins along the minor axis for radial velocity measurements (SNR=4) and 5 spatial bins along one side of the major axis (SNR=7) for velocity dispersion measurements. We can not measure the velocity dispersion along the minor axis due to the 1.8X spectral resolution degradation caused by the large tilt of the slit to the direction of dispersion.
\vskip -4mm
\subsection{Star counts and 2-D photometry} \label{sec:st_counts}
We use the DOLPHOT software package \citep{2000PASP..112.1383D,2016ascl.soft08013D} for crowded field photometry to identify the resolved stars in the HST/ACS \textit{F606W} and \textit{F814W} images and to perform PSF photometry. Only the stars with good quality photometry are included in the final catalogue, following the DOLPHOT recipe and parameters. We choose only the stars lying above the completeness limit $m_{F606W}<27.65$~mag and $m_{F814W}<26.8$~mag.
The spatial distribution of stars is fit with a single-S\'{e}rsic model using maximum likelihood. We set the centre coordinates, axial ratio, positional angle, effective radius, global normalization, and S\'{e}rsic index as free parameters. We also account for foreground contamination following \citep{Makarova10} which gives a total of 30-40 stars from our galaxy. Despite that, the distribution of stars requires an additional constant background component at a level of $\sim 190$ stars which may be an effect of contamination from M~81, M~82, and NGC~3077.

For dynamical modelling we need a global luminosity distribution that is difficult to construct from HST frames with conventional 2D-photometry packages (like Galfit or Sextractor) due to the presence of resolved stars.  We therefore determine the structural parameters of KDG~64 from the magnitude-limited star counts normalized by the surface brightness profile obtained from seeing-limited ground-based images.  We perform a 2-D photometric decomposition of ground-based photometric data using {\sc Galfit} \citep{Peng10}. The model takes centre coordinates, S\'{e}rsic index, effective radius, positional angle, axis ratio and total magnitude as free parameters. All of the fitted parameters except effective radius and PA are close to those obtained from star counts. Comparing our 2D photometry results to the analysis of Subaru data \citep{2019ApJ...884..128O}, we obtain very similar $M_V=-13.37$~mag and the effective radius $R_e=1.1$~kpc. However, the $R_e$ estimate from star counts is slightly smaller at 0.96~kpc. The divergence in $R_e$ and PA might be explained by the presence of a galactic cirrus at the northern edge of KDG~64 that skews the 2-D photometric data in both Subaru and KPNO data. We use the central surface brightness from 2D decomposition as a normalization to convert the surface density profile of the detected stars into surface brightness profile while keeping all the structural parameters from star counts, as this method should better track the mass distribution in KDG~64. 

\begin{figure*}
    \centering
    \begin{minipage}[b]{0.455\linewidth}
    \includegraphics[width=0.95\linewidth]{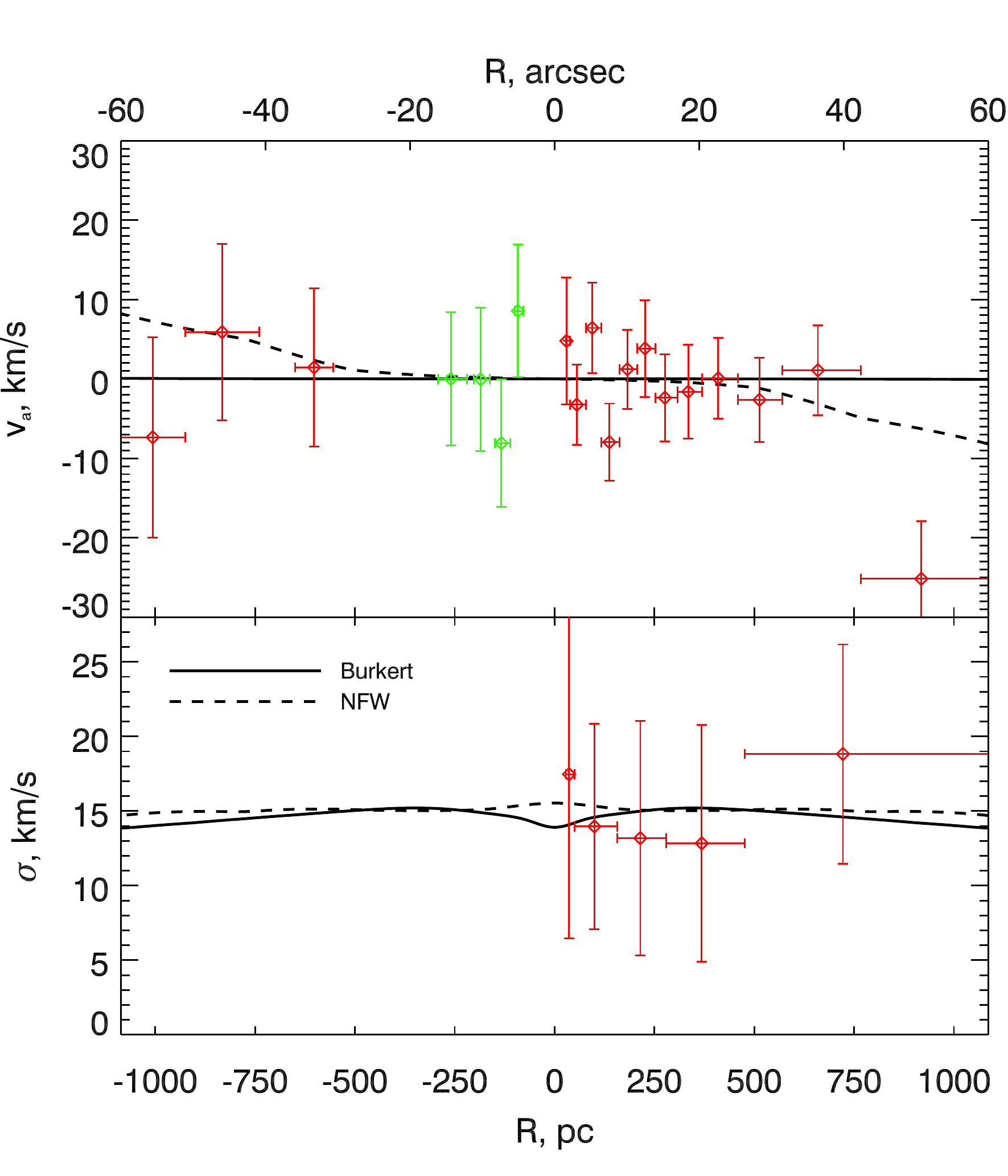}    
    \end{minipage}
    \begin{minipage}[b]{0.37\linewidth}
    \includegraphics[width=0.9\linewidth]{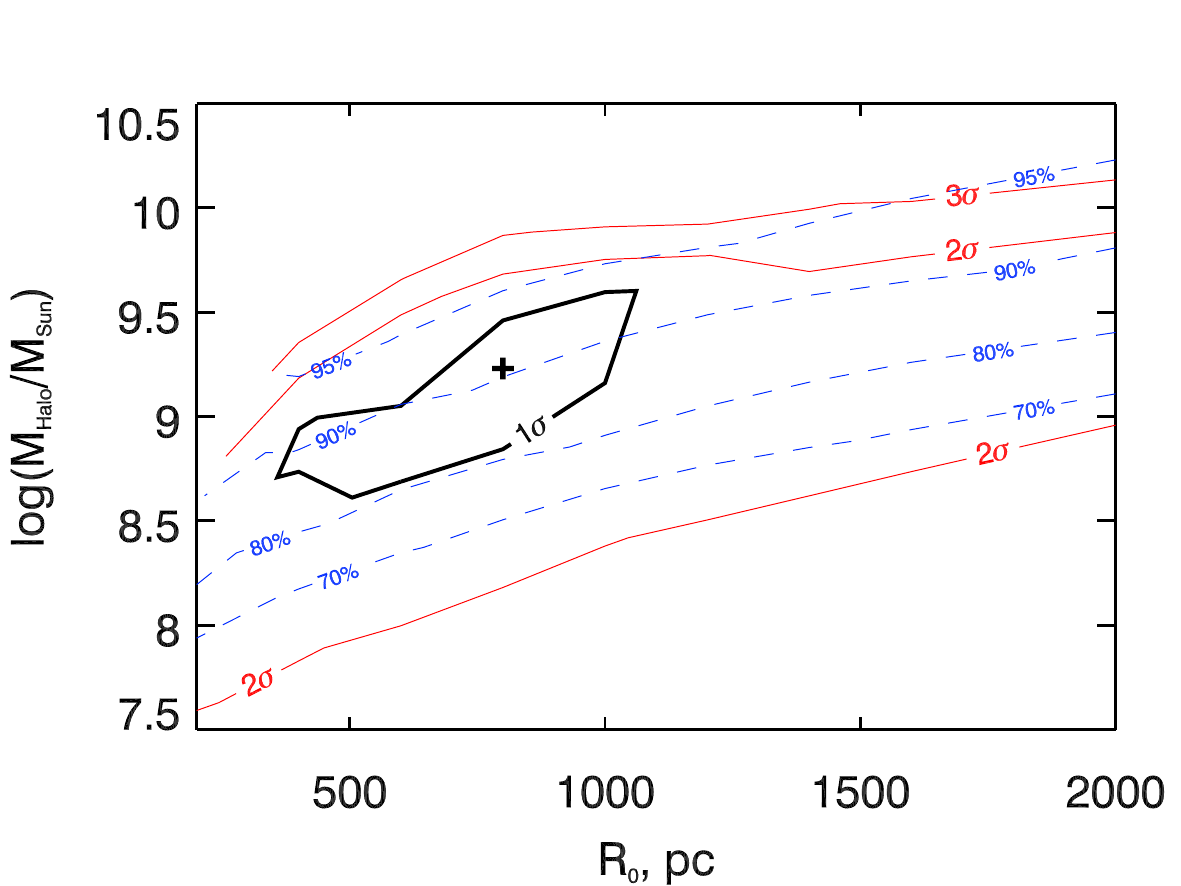}\\    
    \includegraphics[width=0.9\linewidth]{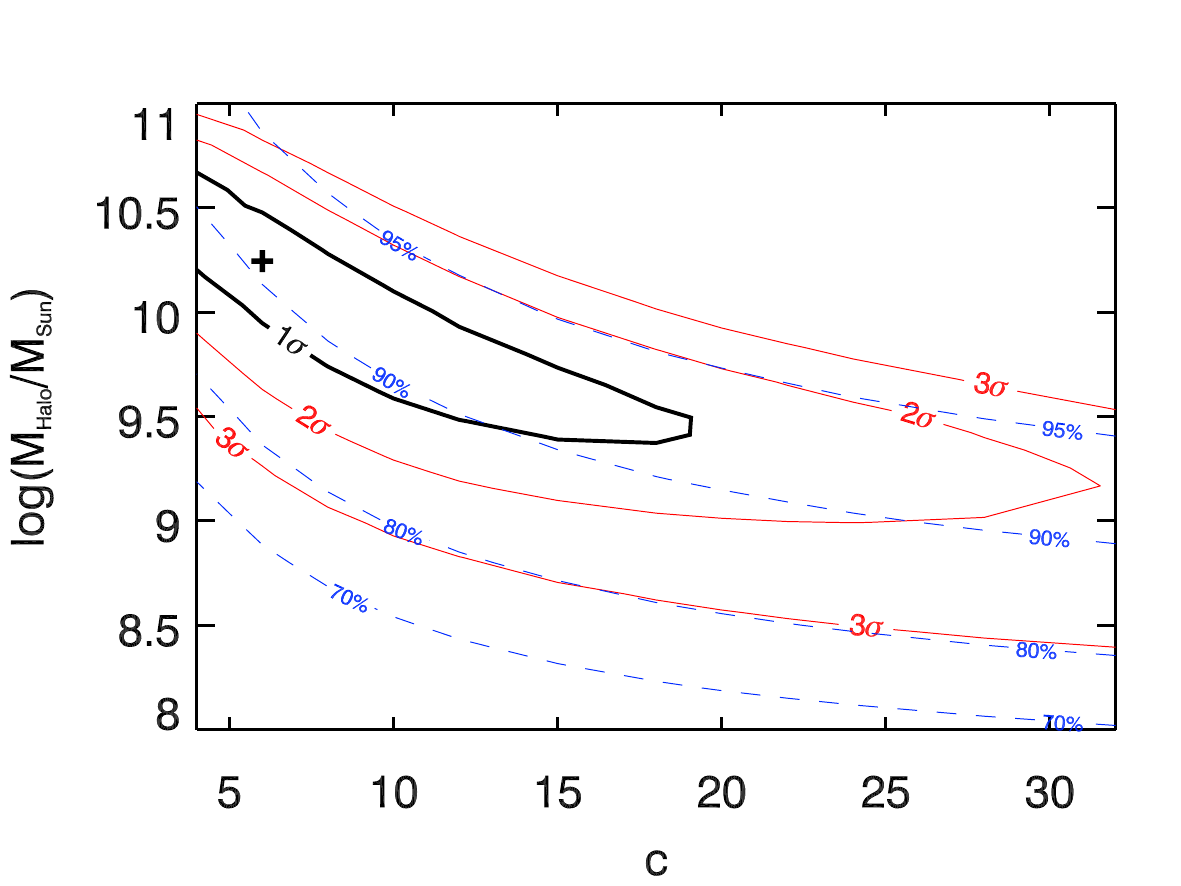}
    \end{minipage}

    \caption{\textbf{Left:} Resolved kinematics profiles (red and green $v_r$ data points for minor and major axes respectively) and best-fitting Jeans dynamical models (black lines) with Burkert and NFW dark matter halo profiles. \textbf{Right:} confidence levels for the parameters of Burkert and NFW DM profiles.}
    \label{fig:dyn_mod}
\end{figure*}

\subsection{Dynamical modelling}
We use the Jeans Anisotropic Modeling (JAM) approach \citep{Cappellari08} to estimate dynamical mass-to-light ratio and mass of dark matter in KDG~64. We follow a procedure described in \citet{2018MNRAS.477.4856A} without the central black hole component. From the $M_{BH} - M_{\mathrm{bulge}}$ scaling relation \citep{kormendy13}, massive central black hole is unlikely to be present in the centre of dwarf spheroidal galaxy (expected black hole mass $< 10^{4.5} M_{\astrosun}$). Currently there is no consensus on the presence of massive black holes in dwarf spheroidals. Most kinematical studies do not find black holes more massive than $10^5~M_{\astrosun}$ \citep[e.g. ][]{2009ApJ...699L.113L}, however there is an evidence that some dwarf spheroidals might host supermassive black holes \citep{2021ApJ...921..107B}.

The basic JAM code does not allow a non-self-consistent (that is ``mass follows light'') as a default, but the code can be easily modified to handle the dark matter halo as a separate component influencing only the overall galaxy potential. We use two dark matter profiles to model KDG~64, a Burkert halo \citep{Burkert95} and a NFW halo \citep{NFW}. These options allow us to model either cored or cusped dark matter distribution and compare the goodness of fit for the kinematic data between the two options. Each of these two DM profiles is described by two free parameters and the other properties can be analytically derived. For the Burkert halo we use $\rho_0$ and $r_s$, and for NFW profile we take $M_{\mathrm{total}}$ and $R_s$. We also calculate the total halo mass for the Burkert profile and the dark matter fraction within the effective radius $R_e$ for both profiles to compare the halo parameters with each other and with the results from \citet{Chilingarian+19}. With this approach we fix the stellar mass-to-light ratio at the value determined from the stellar population, $M/L_{*,R}=1.3~(M/L)_{\odot}$ computed using {\sc pegase.2} models for the stellar population properties of KDG~64.  We model only spherically-symmetric dark matter haloes; oblate or prolate dark matter distributions are outside the scope of our study.

\begin{table}
    \caption{Structural parameters of KDG~64 measured using star counts in HST ACS F814W image (left) and using Galfit analysis of 2.1m KPNO R-band deep images.}
    \label{tab:str_param}
    \centering
    \begin{tabular}{c|cc}
    \hline
    Method & Star counts & 2D decomposition \\
    \hline
    M$_V$, mag & - & -13.37 \\
    $\mu_0$, mag~arcsec$^-2$ & - & 22.77 \\
    $\langle\mu_e\rangle$, mag~arcsec$^-2$ & - & 23.89 \\
    R$_e$,~pc & $960 \pm 25$ & $1125 \pm 20$\\
    R$_e$,~arcsec & $53.1 \pm 1.4$ & $62.2 \pm 1.1$ \\
    n & $1.06 \pm 0.05$ & $1.07 \pm 0.01$ \\
    b/a & $0.45 \pm 0.01$ & $0.49 \pm 0.02$ \\
    PA, deg & $17.4 \pm 1.2$ & $28.1 \pm 0.5$ \\
    \hline
    \end{tabular}
\end{table}

\begin{table}
    \caption{Internal kinematics and stellar population properties of KDG~64 derived from Binospec data.}
    \label{tab:kin_param}
    \centering
    \begin{tabular}{c|cc}
    \hline
    Slit & Major axis & Minor axis \\
    \hline
    S/N, \AA$^{-1}$ & 25 & 10 \\
    $v_{\mathrm{los}}$, km~s$^{-1}$ & $-18.4 \pm 3.4$ & $-18 \pm 6.0$ \\
    $v_{\mathrm{rot}}$, km~s$^{-1}$ & $<2.0$ & $<6.0$\\
    $\sigma$, km~s$^{-1}$ & $16.8 \pm 1.9$ & $17$ (fixed)\\
    Age, Gyr & $10.9 \pm 1.0$ & $11.0$ (fixed) \\
    {[Fe/H]}, dex & $-1.33 \pm 0.26$ & $-1.33$ (fixed) \\
    \hline
    \end{tabular}
\end{table}

The galaxy stellar potential is obtained from the surface brightness profile (see Section~\ref{sec:st_counts}) using the Multiple Gaussian Expansion (MGE) method \citep{2002MNRAS.333..400C}. For dynamical modelling we convert the dark matter profile into multiple Gaussians by fitting 1D dark matter density profiles with the {\sc mge\_fit\_1d} procedure from \citet{2002MNRAS.333..400C} and deprojecting the resulting Gaussians in 3D.  We add the Gaussians describing the dark matter and the stellar population potentials to obtain the final potential. We run JAM over a parameter grid of anisotropy $\beta_z$ (0 to 0.9), and inclination $i$ from edge-on ($90^{\circ}$) to the minimum allowed by the galaxy ellipticity ($64^{\circ}$) in 6$^{\circ}$ steps. The grid also includes two dark matter halo parameters (central density $\rho_0$ and scale radius $r_s$ for Burkert haloes; halo mass $M_{200}$ and concentration $c$ for NFW haloes). We also calculate the halo mass for the Burkert halo (uniquely derived from each pair $\{\rho_0,r_s\}$) for comparison with the NFW halo mass. The concentration $c$ for the NFW profile is capped at $c=4$, because for lower $c$ values it is not always possible to perform a MGE expansion.

\section{Results}
We obtain structural properties of KDG~64 ($R_e$, n, $b/a$) from star counts yielding best-fitting values $R_e=53.1$~arcsec, $n=1.06$, $b/a=0.45$. The light distribution is very close to exponential. Assuming a TRGB-estimated distance modulus of $27.86$~mag \citep{2016AJ....152...50T}, we estimate its physical size as $R_e=960$~pc, placing it among the largest galaxies of the dSph type. For M~81, the adopted TRGB distance modulus is $d=27.79$~mag \citep{2016AJ....152...50T}.  We are able to calculate the full 3D distance between M~81 and KDG~64 of $r=160$~kpc (smaller than the 230~kpc assumed in \citealp{Makarova10}). The KDG~64 and M~81 heliocentric velocities differ by $20$~km~s$^{-1}$ suggesting that the orbit of KDG~64 around M~81 lies nearly in the plane of the sky, in agreement with \citet{Makarova10}. Unfortunately it is difficult to assess if KDG~64 is located closer to the pericentre or the apocentre of its orbit.

\begin{table}
    \caption{Best-fitting parameters of the dynamical models and their 1-$\sigma$ uncertainties.}
    \label{tab:dm_param}
    \centering
    \begin{tabular}{c|cc}
    \hline
    DM Halo & Burkert & NFW \\
    \hline
    log(M$_{200}$/M$_\odot$) & $9.2 ^{+0.3}_{-0.5}$ & $10.2 ^{+0.4}_{-0.7} $\\
    R$_{200}$, kpc & $24 ^{+6}_{-8}$ & $52 \pm 20$ \\
    R$_0$,~kpc or R$_s$,~kpc & $0.8 \pm 0.35$ & $8.8 ^{+9}_{-7} $ \\
    log($\rho_0$), M$_{\odot}~\mathrm{pc}^{-3}$ or c & $-1.0 \pm 0.25$ & $6 ^{+12}_{-2} $ \\
    DM within R$_e$, per~cent & $90 ^{+4}_{-8}$ & $91 ^{+3}_{-4}$ \\
    $\beta_z$ & $0.8 \pm 0.1$ & $0.75 \pm 0.1$ \\
    Inclination, deg & $72 \pm 6$ & $72 \pm 6$ \\
    \hline
    \end{tabular}
\end{table}

The mean stellar age obtained from the SED supplemented full-spectrum fitting is $10$~Gyr, the metallicity is [Fe/H]$=-1.3$~dex, corresponding to a stellar mass to light ratio $M/L_{\mathrm{R,*}}=1.29 \pm 0.11~ (M_{\odot}/L_{\odot,\mathrm{R}})$. The SED clearly indicates the lack of young or intermediate stellar populations with ages 5~Gyr or younger.

The vertical anisotropy $\beta_z$ is rather high, with the best fitting values falling in the range $0.75-0.8$ for both DM haloes; the best-fitting inclination to the line of sight is $i=70$~deg. This means that the shape of KDG~64 is a very oblate spheroid, well represented by a thick disc and not a thin disc geometry.

Cored (Burkert) and cusped (NFW) dark matter distributions fit equally well ($\Delta\chi^2\approx 0.2$), preventing us from probing the shape of the innermost DM distribution. The best-fitting halo total masses differ by an order of magnitude ($\log (M_{\mathrm{Burkert}}/M_{\odot})=9.2$ vs $\log (M_{\mathrm{NFW}}/M_{\odot})=10.2$). However, the dark matter fraction inside the effective radius is the same for both halo profiles, slightly more than 90~per~cent (see iso-lines in Fig.~\ref{fig:dyn_mod} and Table~\ref{tab:dm_param}). 

\section{Discussion}
We demonstrate that we can obtain spectra of low-surface brightness dwarf spheroidal galaxies beyond the Local group of sufficient quality to determine reliable stellar populations and to make spatially resolved kinematic measurements. We are able to perform Jeans dynamical modelling and to estimate dark matter halo parameters. The stellar population parameters of KDG~64 are well measured and consistent between different studies, so we can emphasize kinematics measurements. Here we discuss how KDG~64 can be classified, compare it to other low-luminosity early-type galaxies in the scaling relations, and explore its origin and evolution.

\subsection{Classification of KDG~64}

The first UDG selection criteria were put forward by \citet{2015ApJ...798L..45V} as a result of their observational campaign using a low-resolution Dragonfly telescope array. These criteria only include central surface brightness $\mu_{g,0}>24.0$~mag and the effective radius $R_e>1.5$~kpc, both measured from shallow ($\sim$600~sec) CFHT images with the uniform S\'{e}rsic index $n=1$. Later, \citet{Koda15} used deep Subaru SuprimeCam data to produce a complete catalogue of Coma cluster UDGs and found that to recover all 47 galaxies from \citet{2015ApJ...798L..45V} they had to soften the half-light radius threshold to $R_e>0.7$~kpc. The extrapolated central surface brightness anti-correlates with the assumed S\'ersic index so that for low-luminosity early-type systems having $n<1$ (frequently found in clusters and groups) it will eliminate the brightest objects from the selection. Also, the estimate of the central surface brightness in a galaxy with a central cluster (like one object in the sample of \citet{Chilingarian+19} or several known UDGs in the Virgo cluster) can be misleading because it will not reflect the properties of a galaxy spheroid. \citet{Koda15} used $\langle \mu_{e,R} \rangle$, the mean surface brightness within $R_e$ in the $R$-band, as it represents a less model-dependent metric than $\mu_0$ and is also less sensitive to potential variations of stellar ages compared to the $g$ band. Later, the $r$- and $R$-band mean surface brightness was adopted as a UDG selection criterion by several other teams \citep[see e.g.][]{2018MNRAS.478.2034R,Chilingarian+19,2021NatAs...5.1308G}.
This historical peculiarity has led the smaller intermediate objects with $0.7<R_e<1.5$~kpc being relatively overlooked in studies beyond Local group, while their importance as tracers of the evolution of group and cluster environment is on par with truly large UDGs.

\begin{figure*}
\centering
\includegraphics[width=\hsize]{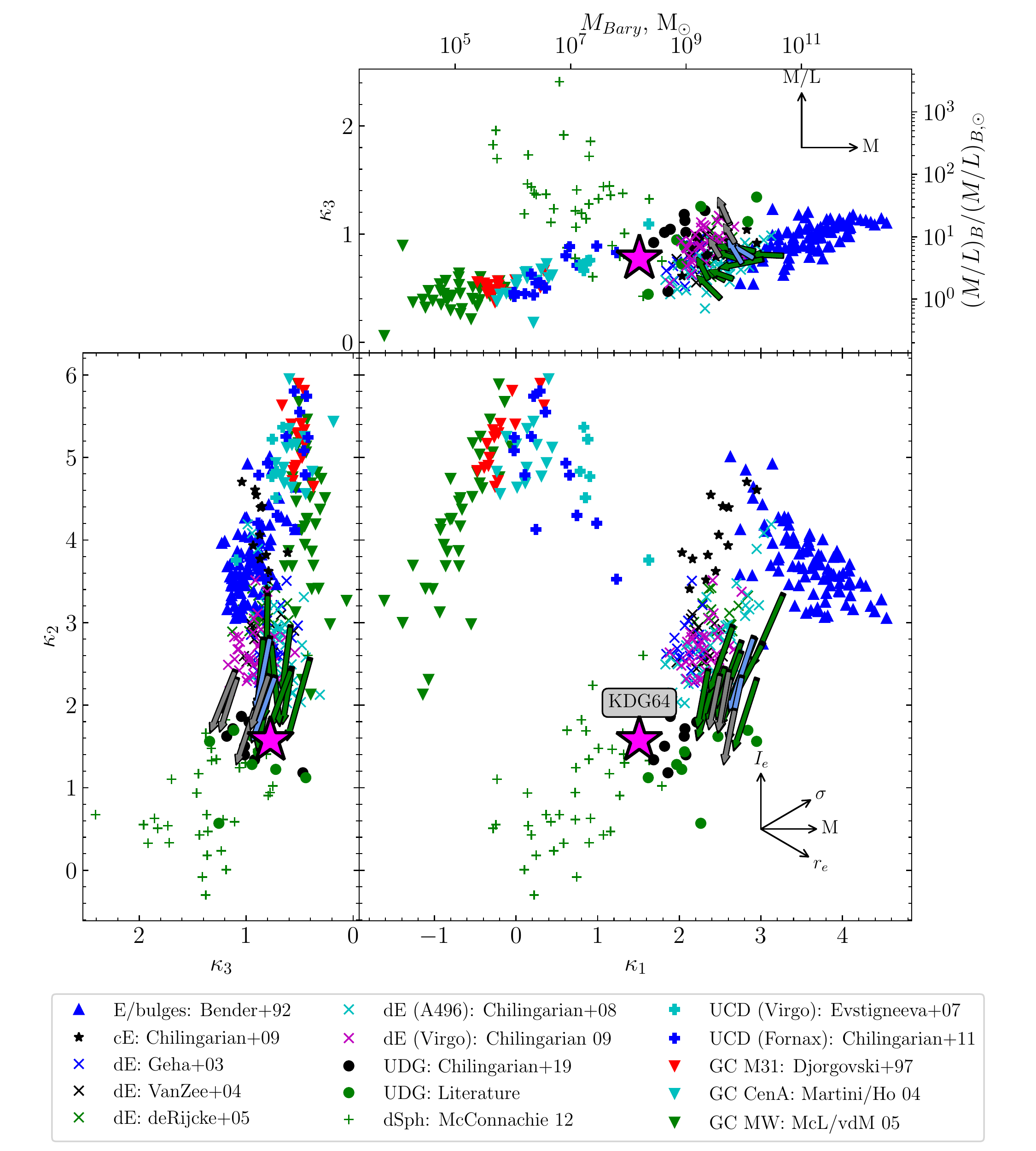}
  \caption{The $\kappa$-space view of the fundamental plane. KDG~64 is marked by a pink star. The sources of data in the literature compilation are shown in the legend; the symbols for different datasets correspond to those in Fig.~\ref{fig:plot_sbrel}. The literature UDG data are from \citet{2017ApJ...844L..11V,Danieli+19,2019MNRAS.484.3425M,2021MNRAS.500.1279F,2020MNRAS.495.2582G,2021MNRAS.502.3144G} and \citet{2023MNRAS.518.3653G}. The arrows indicate the position of dwarf post-starburst UDG and dE progenitors from \citet{2021NatAs...5.1308G} as a result of passive evolution in the next 10~Gyr. The green, blue, and grey arrows correspond to the main Coma sample, main Abell~2147 sample, and an additional sample from that study. Stellar rotation was taken into account where the corresponding measurements were provided \citep{1992ApJ...399..462B,Chilingarian09,2021NatAs...5.1308G}. The baryonic mass calibration of $\kappa_1$ follows \citet{1992ApJ...399..462B}, the $(M/L)$ calibration of $\kappa_3$ was derived from the assumption of axisymmetric S\'ersic profiles.}
  \label{fig:fp}
\end{figure*}

\subsection{KDG~64 in the family of Local Volume dwarf galaxies}

KDG~64 might look similar to the dSph satellites of Milky Way and Andromeda at a first glance, however it is more massive and luminous then any passive dwarf galaxy in the Local Group. The closest local object comparable to KDG~64 is Andromeda VII \citep{2006MNRAS.365.1263M}, but it has slightly smaller $R_e$ (0.75~kpc vs 0.96~kpc) and roughly the same luminosity (both $M_V \sim 13.3$), making it more compact.
The neighbours of KDG~64 in the M~81 group with comparable structural properties are KDG~61, DDO~78, DDO~71 (KDG~63), IKN \citep{2004AJ....127.2031K}, and F8D1 \citep{1998AJ....115..535C} and F12D1. Their $M_V$ range from $-11.5$~mag (IKN) to $-13.9$~mag (KDG~61), and they are on the bright end of the dSph luminosity distribution. These galaxies all have low surface brightness, around $\langle\mu_{e,r}\rangle=24.4$~mag~arcsec$^{-2}$ for KDG~61 and $\langle\mu_{e,r}\rangle\approx25$~mag~arcsec$^{-2}$ for the remaining three galaxies. IKN hosts an unusually rich system of globular clusters \citep{2006AJ....131.1361K} similarly to Fornax dSph \citep{1998ApJ...501L..33B}, but its study is severely hampered by a $9$~mag foreground star projecting directly on the northern part of the galaxy.
\citet{2010A&A...521A..43L} studied stellar populations of all the galaxies mentioned above as well as KDG~64 with archival HST data. They found that the galaxies are generally similar in their metallicities, ages and star formation histories, thus forming a cohesive population of dwarf M~81 satellites.  Future integrated light spectroscopic studies of these objects will help us understand the origin and evolution of UDG and dSph galaxy classes.

The structural parameters of KDG~64 also resemble those of smaller UDGs in the Coma cluster (\citealp{Koda15,yagi16} where the UDG selection criteria were $R_e>0.7$~kpc and $\langle\mu_{e,r}\rangle>24.0$~mag~arcsec$^{-2}$). KDG~64 has $R_e\sim0.95$~kpc and $\langle\mu_{e,r}\rangle=23.89$~mag, so if placed inside Coma cluster and accounting for corresponding cosmological dimming ($\langle\mu_e\rangle$ becomes 24.01), KDG~64 would have barely passed the \citet{yagi16} UDG selection criteria. KDG~64 has a rather elongated shape, so if deprojected, its surface brightness would be even fainter. \citet{yagi16} catalog contains 204 galaxies (out of total 854) with $R_e$ smaller than KDG~64, and 101 of them have $b/a<0.5$.  From a morphological perspective KDG~64 is similar to the small UDGs in the Coma cluster. These three similarities motivate the claim that KDG~64 is one of the closest UDG analogs in the Local Universe. It could act as a calibration object for integrated light studies of more distant galaxies.

\subsection{KDG~64 kinematics}
KDG~64 is located $97.5~\mathrm{arcmin} = 106$~kpc in projected distance from M~81. Even without taking into account the line of sight distance difference, we can estimate the maximum tidal disturbance from M~81. \citet{2002A&A...383..125K} derive M~81's group total dynamical mass as $\sim 1.6\times10^{12} M_{\odot}$, so the mass of the M~81 dark matter halo can be estimated as about 2 times as small, in agreement with \citet{Oehm17}. An initially spherical galaxy $106$~kpc from M~81 would have been distorted into an ellipsoid with expected axis ratio $a/b=1.16$ in the M~81 direction. 
This value is much smaller than $a/b=2.2$ we obtain from the KDG~64 photometric analysis. Additionally, KDG~64's isophotes are extended along the line perpendicular to the direction towards M~81. It is therefore unlikely that KDG~64's elliptical shape stems from its current tidal interaction with M~81.

Using the 3D distance $d=160$~kpc we can constrain the sphere of KDG~64's gravitational influence from the Jacobi radius (see \citealp{Binney+Tremaine2}):
\begin{equation}
R_j=d\times\left( \frac{M_\mathrm{KDG~64}}{3M_\mathrm{M81}}\right)^{1/3}=24~\mathrm{kpc}.
\end{equation} 
Interestingly, the best-fitting value for $R_{200}$ of the Burkert dark matter halo is $\sim 25$~kpc suggesting that the dark matter haloes of KDG~64 and M~81 are in equilibrium. Thus, if KDG~64's dark matter halo is being stripped by its host, the process is gradual and should not lead to halo truncation \citep[see also][]{2022MNRAS.509.5330B}. 

KDG~64 exhibits no rotation along the major axis up to $20$ arcsec ($\sim 360$~pc, or $1/3$~R$_e$), and beyond that radius the presence of the rotation is questionable. The minor axis also does not exhibit rotation, consistent with our initial assumption of axial symmetry. The dispersion profile is flat, suggesting the dark matter halo dominates the potential at all radii. Low rotation and strong dispersion support is typical of dwarf spheroidals \citep{2009ApJ...704.1274W} as well as UDGs (\citealp{Chilingarian+19}, \citealp{2018MNRAS.478.2034R}, \citealp{2019ApJ...880...91V}). 

The $\kappa$-space Fundamental plane \citep{1992ApJ...399..462B} is an important kinematics metric for a dispersion supported virialized galaxy. This is a modification of the original Fundamental Plane \citep{1987ApJ...313...59D} with the rotated axes (see Figure~\ref{fig:fp}). \nocite{2005A&A...438..491D} \nocite{1992ApJ...399..462B} \nocite{2004AJ....128.2797V} \nocite{2003AJ....126.1794G} \nocite{Chilingarian+08} \nocite{2005ApJS..161..304M} \nocite{1997ApJ...474L..19D} \nocite{2004ApJ...610..233M} \nocite{2012AJ....144....4M} \nocite{Chilingarian+09} \nocite{Chilingarian+19} \nocite{2007AJ....133.1722E} \nocite{Chilingarian+11} \nocite{Chilingarian09} The $\kappa$-parameters are defined as: $\kappa_1=(\mathrm{log}\sigma_0+\mathrm{log}R_e)/\sqrt{2}$ (a measure of $M_{\mathrm{dyn}}$); $\kappa_2=(\mathrm{log}\sigma_0+2\mathrm{log}I_e-\mathrm{log}R_e)/\sqrt{6}$ (compactness); $\kappa_3=(\mathrm{log}\sigma_0-\mathrm{log}I_e-\mathrm{log}R_e)/\sqrt{3}$ (a measure of $(M/L)_{\mathrm{dyn}}$).The arrows in Fig.~\ref{fig:fp} indicate the results of passive evolution of 16 diffuse post-starburst galaxies in the Coma and Abell~2147 clusters from \citet{2021NatAs...5.1308G} in the next 10~Gyr. The color indicates one of the three sub-samples: blue for the main sample of 11 Coma cluster galaxies, green for the main sample of 2 Abell~2147 galaxies, and red for the 5 additional objects in both clusters which fall below the SDSS spectroscopic magnitude limit. Two out of sixteen galaxies will passively evolve into dE/dS0s while the remaining 14 will end up in the UDG locus.
The $\kappa_1-\kappa_2$ and $\kappa_2-\kappa_3$ projections show KDG~64 directly at the interface between loci of UDGs (black circles) and `classical' dSphs (green stars). KDG~64's position in the $\kappa_1-\kappa_3$ projection skews to UDG locus, lying in the narrow cloud occupied by GCs, cEs, UDGs, dEs, and giant ellipticals, while most of dSphs are noticeably above this cloud. These metrics clearly display KDG~64's transitional nature between dSphs and UDGs, favoring the UDG locus.
The dark matter fraction within $1 R_e$ of just above 90~per~cent is comparable to the values found in \citet{Chilingarian+19} for small Coma UDGs, and is smaller than for most extended Local Group dwarfs. \citet{2009ApJ...704.1274W} show that the dark matter fraction in dwarf spheroidal galaxies is stable to the choice of dark matter profile, and we demonstrate that it is also true for KDG~64. The 800~pc core size is similar to that of large dSphs (e.g \citealp{2013MNRAS.429L..89A}). 

\subsection{KDG~64 evolution}
The HST ACS data shows no detectable young stellar population. Our full-spectrum+SED fitting results, and especially the GALEX FUV and NUV fluxes support this conclusion. \citet{Makarova10} found a slight enhancement of the star formation mostly about 1.5 $-$ 2.5 Gyr ago and an indication of very small fraction (less than 2~per-cent) of stars $\sim$500 Myr old. These intermediate age stellar populations account for about 10~per~cent of the total mass. \citet{2010ApJ...724.1030G} reports 7 to 9~per~cent of population younger than 3~Gyr (with less than 1~per~cent younger than 1~Gyr) based on the AGB to RGB star ratio. \citet{2011ApJ...739....5W} indicate a more protracted star formation history, with only a half of KDG~64 stellar mass in place 6~Gyr ago, and $18$~per~cent of the stellar mass formed during the last 2~Gyr. This result is inconsistent with our results from full-spectrum fitting and SED analysis. \citet{2010A&A...521A..43L} do not infer age, giving only the metallicity which varies between $-1.72$~dex and $-1.39$~dex depending on the chosen normalization and isochrone. CMD-based results from Milky Way dSphs \citep{2009ARA&A..47..371T} suggest that many of them went through multiple phases of moderate star formation after the buildup of the bulk of their stellar mass $\sim 10$~Gyr ago. These events typically make up about 10 per cent of the stellar mass, and happen once in 3-5 Gyr. 

The best-fitting model for the stellar distribution was found to be S\`{e}rsic~+~constant background. This background could be attributed to either M~81~+~M~82~+~NGC~3077 peripheral stars, or excess foreground stars in our Galaxy. In both cases, this background could contain some young stars that might shift the age statistic derived from CMD slightly towards younger ages, and the number of young stars detected in \citet{Makarova10} is generally comparable to this background value. 

\citet{2001ApJ...560L.127B} have detected an H{\sc i} spur extending from the main cloud around NGC~3077 to KDG~64. However it is mostly located at the radial velocity of NGC~3077, which is about 200~kpc farther than KDG~64. This association with H{\sc i} is most likely a projection effect, hence KDG~64 does not have any neutral gas.

The evolutionary path of dwarf spheroidal galaxies is still a matter of debate as no single mechanism explains the diversity of their properties \citep{2001ApJ...547L.123M,2007Natur.445..738M,2005MNRAS.356..107R}. Here we use KDG~64 properties as a case study for the main dSph formation theories. 
AGN in dSph have not been discovered to date, but according to well-established scaling between central black holes and their host galaxies \citep{2000ApJ...539L...9F,2000ApJ...539L..13G,2004ApJ...604L..89H}, they are expected to be in the intermediate-mass range ($2$ to $5 \times 10^4 M_{\odot}$), which will not provide enough energy output to quench star formation in the entire galaxy even if the AGN accretes at the Eddington limit.
The starvation and strangulation mechanisms are also unlikely to play a major role, for their ineffectiveness in the low mass regime allows dwarf galaxies to form stars in the absence of an external gas reservoir for time periods comparable to Hubble time.

Supernovae (SN) feedback \citep{1986ApJ...303...39D} was widely considered the most common quenching mechanism in the low-mass stellar systems. In the absence of external forces, SN feedback is theorized to ``fire'' only once in the low mass galaxies, as the gas ejected by SN remnants is expelled forever from a galaxy potential far into the IGM.  However, the typical star formation rate in a galaxy with similar mass to KDG~64 does not generate enough SNe to expel all the gas. Many similarly massive galaxies (i.e. from DDO catalogue) are still star-forming, meaning SN feedback was not effective there. While SN feedback provides a good explanation for the old and metal poor stellar population, it doesn't explain the lack of rotation in KDG~64. Dynamical modelling suggests KDG~64 is a very oblate spheroid that does not rotate. However, if it was formed from a star forming discy dIrr progenitor via SN feedback, some residual rotation should have been preserved after the quenching.

X-ray data from the XMM telescope archive\footnote{\url{https://nxsa.esac.esa.int/nxsa-web/search}} do not demonstrate that M~81 has a massive halo of hot circumgalactic gas. It is well established that ram-pressure quenching is not as efficient in groups as in clusters, however it still might be enough for a $2\times10^7~M_{\odot}$ galaxy. \citet{2019A&A...624A..11H} show that dense pressurized gas experiencing ram-pressure might actually shield itself from evaporation, caused either by ram-pressure or UV background during reionization epoch. However, some models show that in the group environment ram pressure could act together with tidal heating to completely quench a dwarf spheroidal satellite \citep{2006MNRAS.369.1021M}.

It is also puzzling that KDG~64 does not contain globular clusters (GCs) or a central star cluster (the central object visible in Fig.~\ref{fig:mask} is a background galaxy). Some morphologically similar galaxies contain many globular clusters (Fornax, IKN), while some do not contain any (Andromeda VII, KDG~61) and over half of all dwarf elliptical galaxies are nucleated \citep{2005MNRAS.363.1019G}. The majority of Milky Way dSphs do not have any GCs, however they are typically less luminous than KDG~64. Evidently, there exists some mechanism to remove them completely, and SN feedback is not expected to influence globular clusters. 

We suspect that ram-pressure stripping, SN feedback and disc heating by M~81 tides all could have played a role in KDG~64 evolution. The old homogeneous stellar population and high orbital anisotropy suggest that KDG~64 should have spent most of its lifetime at moderate distance from M~81, where tidal forces are not strong enough to significantly deform the dSph satellite or strip its DM halo, but where tidal forces significantly contribute to the kinematical heating of the stellar component. As a result the stars gradually shift to the eccentric orbits, and the galaxy transitions from being rotation-supported to pressure-supported.  Globular clusters are also shown to be influenced by tides that puff up the GC distribution \citep{2021MNRAS.502..398C}, easing their subsequent detachment. It is possible that the galaxy is initially quenched by SN feedback, and starts to be tidally heated soon afterwards. This tidal evolution has been shown to be important in the models of the Local Group dSphs \citep{2008ApJ...673..226P}.

KDG~64 demonstrates that large group-dwelling dwarf spheroidal galaxies are in fact barely distinguishable from other diffuse low-mass non-starforming stellar systems in groups and clusters.  It is not possible to make sweeping conclusions from a single galaxy, but we can argue that at least some of the morphologically similar cluster dwarfs should follow the same evolutionary paths as dwarf spheroidals in groups. The definitive answer requires systematic observations of low mass galaxies in different environments analysed in a homogeneous fashion.  This work demonstrates the feasibility of such a program.

\section*{Acknowledgements}

AA acknowledges the Universit\'e de Paris for funding his PhD research and the Russian Science Foundation (RScF) grant No. 22-12-00080 for supporting the development of the dynamical modelling procedures. IC, KG acknowledge the RScF grant No. 19-12-00281 for supporting the analysis of the spectral and photometric data and the Interdisciplinary Scientific and Educational School of Moscow University ``Fundamental and Applied Space Research''. We thank the anonymous referee for valuable comments, which helped us to improve the manuscript. IC's research is supported by the Telescope Data Center at Smithsonian Astrophysical Observatory. We are grateful to the staff of the MMT Observatory jointly operated by Smithsonian Astrophysical Observatory and the University of Arizona for their support of Binospec operations and service mode observations. We thank F.~Combes, G.~Mamon, and O.~Sil'chenko for fruitful discussions related to this project.

\section*{Data Availability}
The data underlying this article will be shared on reasonable request to the corresponding author.
%%%%%%%%%%%%%%%%%%%%%%%%%%%%%%%%%%%%%%%%%%%%%%%%%%



\bibliographystyle{mnras}
\bibliography{K64.bib} % if your bibtex file is called example.bib


%%%%%%%%%%%%%%%%%%%%%%%%%%%%%%%%%%%%%%%%%%%%%%%%%%

%%%%%%%%%%%%%%%%% APPENDICES %%%%%%%%%%%%%%%%%%%%%

%\appendix


%Extra materials, you want it? It is yours my friend, as long as you have enough persistence.

%%%%%%%%%%%%%%%%%%%%%%%%%%%%%%%%%%%%%%%%%%%%%%%%%%

% Don't change these lines
\bsp	% typesetting comment
\label{lastpage}

\end{document}